\documentclass{article}
\usepackage{spconf,amsmath,graphicx,amssymb}
\usepackage{array}

\usepackage{hyperref}
\usepackage{wrapfig}
\usepackage{graphicx}
\usepackage{subfigure}
\usepackage{float}
\usepackage{adjustbox}
\usepackage[justification=centering]{caption}
\usepackage{tikz}
\usepackage{tabularx}
\usepackage{multirow}
\usepackage[belowskip=-15pt,aboveskip=0pt]{caption}

\newcolumntype{P}[1]{>{\centering\arraybackslash}p{#1}}
\newcolumntype{M}[1]{>{\centering\arraybackslash}m{#1}}
\newcommand{\PreserveBackslash}[1]{\let\temp=\\#1\let\\=\temp}
\newcolumntype{R}[1]{>{\PreserveBackslash\raggedleft}p{#1}}
\usepackage{tabulary,booktabs}
\usepackage{enumitem}
\usepackage{lipsum}
\usepackage{blindtext}
\usepackage[linesnumbered, algo2e, ruled,norelsize]{algorithm2e}
\usepackage{xcolor}
\usepackage{cite}

\newcommand{\Rclr}{R\textsubscript{clr}}
\newcommand{\Tenc}{T\textsubscript{enc}}

\SetCommentSty{mycommfont}
\title{Deep probabilistic model for lossless scalable point cloud attribute compression}
\name{Dat Thanh Nguyen, Kamal Gopikrishnan Nambiar and Andr\'e Kaup \thanks{This work was funded by the Deutsche Forschungsgemeinschaft (DFG, German Research Foundation) under Grant SFB 1483 – Project-ID 442419336.}}
\address{Chair of Multimedia Communications and Signal Processing \\ Friedrich-Alexander-Universität Erlangen-Nürnberg (FAU) \\ Erlangen, Germany}

\begin{document}
%
\maketitle
\begin{abstract}

In recent years, several point cloud geometry compression methods that utilize advanced deep learning techniques have been proposed , but there are limited works on attribute compression, especially lossless compression. In this work, we build an end-to-end multiscale point cloud attribute coding method (MNeT) that progressively projects the attributes onto multiscale latent spaces. The multiscale architecture provides an accurate context for the attribute probability modeling and thus minimizes the coding bitrate with a single network prediction. Besides, our method allows scalable coding that lower quality versions can be easily extracted from the losslessly compressed bitstream. We validate our method on a set of point clouds from MVUB and MPEG and show that our method outperforms recently proposed methods and on par with the latest G-PCC version 14. Besides, our coding time is substantially faster than G-PCC. 


\end{abstract}
\section{Introduction}
\label{sec:intro}
\par A point cloud is a collection of 3D points with spatial coordinates and associated attributes (e.g., color, velocity, etc.). Typical PCs contain millions of 3D points which require a huge amount of storage. Especially for dense point clouds, the color components require a much higher bitrate than the geometry components. As a result, efficient point cloud attribute compression methods are highly desired. Thus, in this paper, we focus on attribute compression for dense point clouds.
\par The Moving Picture Expert Group (MPEG) has been developing two approaches for Point Cloud Compression (PCC) \cite{graziosi2020overview}: Geometry-based PCC (G-PCC) and Video-based PCC (V-PCC). G-PCC directly processes and compresses the point clouds in 3D space while V-PCC projects the point clouds to 2D planes and utilizes 2D advances in image and video compression. In this paper, we assump that the geometry is losslessly encoded and only focus on point cloud attribute compression. 
\par Similar to 2D image and video content, we can exploit the spatial and temporal correlation within the point cloud attribute for compression tasks. However, it is not straightforward to bring the efficient signal processing tools in 2D space onto sparse and unevenly sampled 3D point data. Most existing approaches for attribute compression are based on geometry-dependent transforms \cite{7025414,7482691,gu20193d}, the geometry representation thus plays a vital role in attribute compression. Point cloud geometry can be represented in either point, octree, or voxel representation. Voxel representation can be processed by conventional operations such as 3DCNN, DCT, etc, however, this requires significant computation power. Octree representation has a lower uncompressed bitrate than voxel but octree is not compatible with conventional operations. In this work, we represent point clouds in the point-based representation and utilize the sparse convolution \cite{choy20194d} to efficiently process the geometry and attribute. 
\par To encode the point cloud attributes in a lossless manner, we need to build an accurate context model to estimate the attribute probabilities for a context arithmetic coder. We introduce a multiscale deep context model to estimate the attribute probabilities in which the higher scales are conditioned on lower scales. We demonstrate experimentally that our approach outperforms recently proposed methods in attribute compression and rivals the latest G-PCC version while producing superior computing. The rest of the paper is structured as follows: Section II reviews related work; MNeT method is described in Section III; Section IV presents the experimental results, and conclusions are drawn in Section V.



\section{Related work}
\label{sec:stateoftheart}
%

Many point cloud attribute compression methods rely on geometry-dependent transforms \cite{7025414,7482691,gu20193d}. For instance, the region-adaptive hierarchical transform (RAHT) \cite{graziosi2020overview} transforms the attributes of the occupied octree nodes and then encodes the quantized transform coefficients using the run-length Golomb-Rice encoding (RLGR). The method is later adopted by MPEG G-PCC and becomes one of the core transform modules in the TMC13 reference software \cite{TMC13} besides the lifting transform \cite{lifting} and the predicting transform \cite{predicting}. The proposed method in \cite{7025414} constructs a graph from geometry and treats the attribute as the signal on this graph. The attributes are then transformed using a Graph Fourier Transform (GFT) before being quantized and entropy coded. 
\par One of the directions for point cloud attribute coding is projecting the 3D space onto 2D planes and deploying existing image and video compression standards (e.g., JPEG and H.265). For example, the method proposed in \cite{7434610} partitions the point cloud into blocks of size 8x8 and uses snake scanning before converting to 2D images and encoding by JPEG standard. MPEG V-PCC \cite{graziosi2020overview} projects the point cloud data into a set of three associated videos: occupancy map, geometry video, and attribute video. All three generated videos are then encoded by 2D video coding standards. 

\par Recently, several studies \cite{quach2020folding,9447226,wang2022sparse} have been proposed to apply advanced deep learning techniques to lossy point cloud attribute compression. A folding network to project 3D attribute signals to 2D planes is proposed in \cite{quach2020folding}. The authors in \cite{wang2022sparse} proposed an auto-encoder neural network to map attributes to a latent space and then encode the latent vector using a hyper prior/auto-regressive context model. The point clouds are represented using sparse tensors and can be efficiently processed by a sparse convolution. The auto-encoder architecture is also utilized in the end-to-end lossy attribute coding proposed in \cite{9447226}. Inspired by recent works \cite{9447226, mentzer2019practical} using autoencoder and hyper priors for compression, we build an autoencoder-based multiscale architecture for lossless attribute coding.
\par In fact, most studies in the lossless attribute coding are based on conventional methods \cite{huang2021hierarchical,yin2021lossless,song2021layer} including G-PCC from MPEG \cite{graziosi2020overview}. For example, a hierarchical bit-wise differential coding-based method (HBDC) was recently proposed in \cite{huang2021hierarchical}. The authors conduct an octree breath-first traversal and only encode the residual between parent and child nodes when necessary. In \cite{yin2021lossless}, authors exploit local dependencies by introducing normals to improve the Prediction Transform of G-PCC. However, most of the methods are still inferior to the conventional G-PCC. MPEG G-PCC version 11 was introduced in September 2020, and was recently improved with many efficient tools for attribute coding such as residual coding, attribute prediction scheme, or coordinate conversion and the latest G-PCC version 14 reports the state-of-the-art attribute compression performance. In this paper, we build an efficient learning-based context model for entropy coders using a sparse convolution neural network. To the best of our knowledge, this is one of the very first works on learning-based lossless coding for point cloud attributes. 

\section{Proposed method}
\label{proposedmethod}

\subsection{Context Model}
\begin{figure}
\centering
\captionsetup{justification=justified}
\includegraphics[width=.95\linewidth]{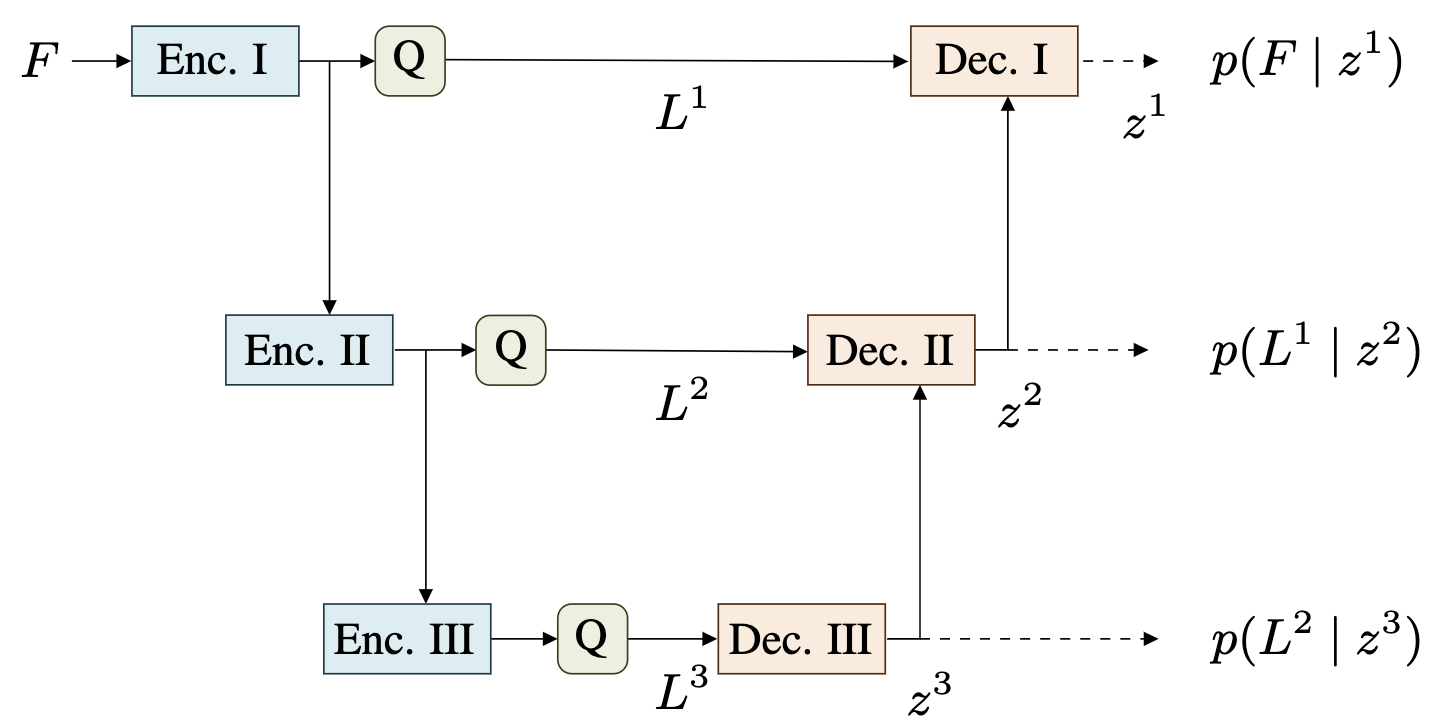}
\vspace{0.1cm}
\caption{System overview of the proposed method MNeT. At each scale, we have three components: Encoder (blue), a Quantizer (green), and a Decoder (light grown). All latent representation (L) and RGB feature (F) are encoded to bitstream using the predicted probability distribution from the decoders, except the last layer where we encode $L^3$ using a uniform distribution.   }
\label{fig:3Dcontext}
\end{figure}

\par The details of our novel method MNeT will be presented in this section. We represent point cloud attributes using spare tensors \{C, F\} given that the coordinates of occupied voxels  $C=\{x_i,y_i,z_i\}_i$ are already known. We aim to losslessly encode the feature set $F$ that is composed of red, green, and blue (RGB) colors, or other attributes such as YUV colors, normals, velocities, etc. Our method losslessly encodes the feature $F$ using context-based arithmetic coding. Specifically, we aim to build a probability model that accurately estimates the joint probability $p(F)$ by introducing hierarchycal auxiliary latent representations $L^1, L^2,.., L^{N}$. The overview architecture of our novel method MNeT is shown in Figure  \ref{fig:3Dcontext}. To encode the attributes, we feed the point cloud sparse tensor through $N$ auto-encoders and extract the latent representation $L^n$ from the encoders. The probability distribution of the feature $F$ and latent vector $L^{1,..,N}$ are predicted by the decoders.  Hence, the joint probabilistic distribution of the feature $F$ and the latent representation $L^{1,...,N}$ is modeled as:

\begin{equation} \label{eq:factorization}
\begin{split}
   p(F, L^{1,..,N})=p(F|L^{1,..,N}) \cdot \prod_{n=1}^{N} &p(L^n|L^{n+1}, .., L^N) 
\end{split}
\end{equation}

\begin{figure*}
\centering
\captionsetup{justification=justified}
\includegraphics[width=.83\linewidth]{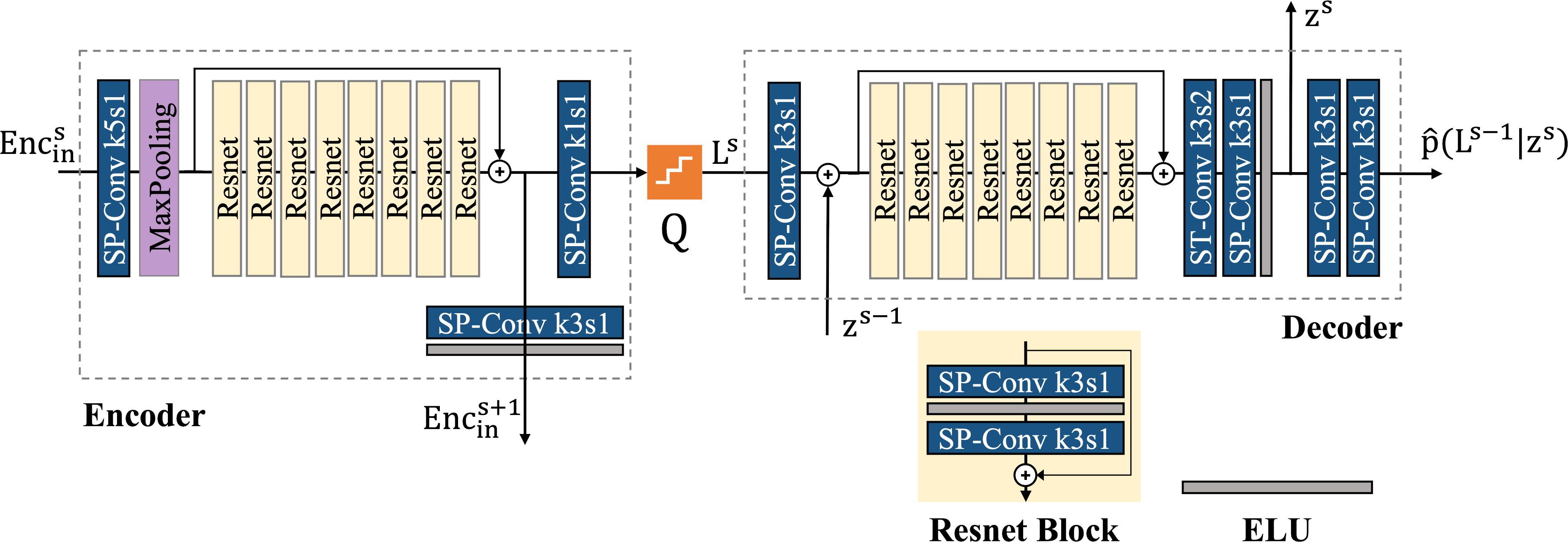}
\vspace{0.1cm}
\caption{MNeT network architecture for a single scale including an \textbf{Encoder}, a Quantizer \textbf{Q} and a \textbf{Decoder}. SP-ConV and ST-Conv denote the sparse convolution and sparse transpose convolution with convolution kernel and stride denoted using kKsS (kernel K, stride S). We use the same number of output filters (filter=64) for all convolution layers, except the one before quantization (filter=5) and the last convolution layer at the decoder (filter $\in$ [120, 150, 150]).}
\label{fig:network}
\end{figure*}

Each term on the right-hand side is the joint probability of latent representation $L^n$ or the RGB color $F$ given all lower scales. However, instead of directly modeling the upper latent representation from the lower scales $p( L^n|  L^{n+1},..., L^N)$, the decoders are modeled such that
\begin{equation} \label{eq:factorization}
\begin{split}
   p( L^n|  L^{n+1},..., L^N)
   &=p(L^n|z^{n+1})
\end{split}
\end{equation}
with $z^{n+1}$ is considered as a sumerization of $L^{n+1,..,N}$. Auto-regressive modeling has been shown as the state-of-the-art method for estimating data likelihood \cite{salimans2017pixelcnn++,nguyen2022learning} by learning the dependencies between sub-points/pixels. However, in this work, we do not factorize the probability $p(F)$ over sub-points, which can lead to a significant number of network evaluations. Instead, we assume that all sub-points are statistically independent and jointly model all the sub-points together while keeping the relation between RGB channels within each sub-point. Specifically, for each sub-point feature $f_i$ of F, we factorize the joint distribution as: 
\begin{equation} \label{eq:factorization}
\begin{split}
   p(f_i|z^1)&=p(f_{i}^{RGB}|z^1)\\
   &=p(f_{i}^{R}|z^1) \cdot p(f_{i}^{G}|f_{i}^{R},z^1) \cdot p(f_{i}^{B}|f_{i}^{R},f_{i}^{G},z^1)
\end{split}
\end{equation}
Each probability distribution on the right hand side is modeled via discretized logistic mixture model proposed in \cite{salimans2017pixelcnn++}. Specifically, we first model the distribution of the red channel using information from the lower latent as a mixture of $K$ logistic distributions. Next, we predict the green channel using the same principle, but we allow the means of mixture components to linearly depend on the value of the red channel. Similarly, in the blue channel, the mean of the mixture coefficients depends on the values of the red and green channels. On the lower scales, we also use a logistic mixture to model the latent distributions, but we do not enforce any dependencies between channels. 

\par In order to learn good predictions $\hat{p}(F, L^1,..., L^N)$  of the joint color attribute and latent distribution $p(F, L^1,..., L^N)$ and thus minimize the encoding bitrate, we train the auto-encoders by minimizing the total cross-entropy losses on all latent representations: 
\begin{equation} \label{eq:lossfunction}
\begin{split}
   Loss &= H(\hat{p},p)=H(\hat{p}_{F|z^1},p_{F|z^1}) + H(\hat{p}_{L^1|z^2},p_{L^1|z^2})  \\
   &+H(\hat{p}_{L^2|z^3},p_{L^2|z^3}) +H(\hat{p}_{L^3},p_{L^3}) 
\end{split}
\end{equation}

\subsection{Network Architecture}
\par The architecture of an auto-encoder for a single scale is presented in Figure \ref{fig:network}. At the encoder side, we first apply a convolution followed by a Max-Pooling layer. We find that aggregating a convolution layer with a Max-Pooling provides faster training and better performance compared with a single convolution layer with stride=2. Next, we deploy 8 residual blocks followed by two convolution layers in two branches: one going to the quantization (filter=5) and one going to the lower scale (filter=64). The MNeT residual blocks are inspired by the 2D residual block from EDSR \cite{lim2017enhanced} as our multi-scale prediction is similar to super-resolution tasks. We adopt the scalar quantization from \cite{mentzer2019practical} which allows gradient back propagation during training. We set the number of bins to 26 in all the quantizers. At the decoder side, which takes inputs from the quantization and the lower scale, we first apply a convolution layer followed by 8 residual blocks, an upsampler, and convolutions. The decoders output the logistic mixture parameters for probability construction and the forwarding feature for the upper scale.  In this work, we use 10 mixture of logistics on every scale, thus the number of decoders' output channels are 120, 150, 150 from the top scale to the bottom scale, respectively.

\begin{table*}[t]
\caption{Average rate in bpp and encoding time of the proposed method compared with HBDC and MPEG G-PCC}
\label{tab:compare}
\begin{center}
\resizebox{0.9\linewidth}{!}{
\begin{tabular}{@{}lccccccccccccc@{}}
	\toprule
	&&\multicolumn{2}{c}{\textbf{G-PCC v14}}
	&\phantom{--}&\multicolumn{2}{c}{\textbf{G-PCC v11}}
	&\phantom{--}&\multicolumn{2}{c}{\textbf{HBDC}}
	&\phantom{--}&\multicolumn{2}{c}{\textbf{MNeT}}\\
	\cmidrule{3-4} \cmidrule{6-7} \cmidrule{9-10} \cmidrule{12-13}
	
	\textbf{Point Clouds} & \textbf{No. points}     
	&\textbf{\Rclr} &\textbf{\Tenc} 
	&&\textbf{\Rclr} &\textbf{\Tenc} 
	 &&\textbf{\Rclr} &\textbf{\Tenc} 
	&&\textbf{\Rclr} &\textbf{\Tenc}\\

	&&(bpp)&(s)&&(bpp)&(s)&&(bpp)&(s)&&(bpp)&(s)\\

	\midrule
Phil&1559008&10.25&6.99&&14.12&13.67&&12.67&0.44&&10.49&3.89\\
Ricardo&956482&6.03&3.76&&8.13&8.50&&7.91&0.29&&7.31&3.15\\
Andrew&1286783&11.19&10.90&&15.19&10.90&&12.91&0.37&&12.27&3.80\\
Redandblack&693899&9.64&2.97&&10.27&5.91&&11.39&1.13&&12.23&2.68\\
Loot &784142&6.57&3.29&&8.89&6.46&&7.79&1.49&&8.38&3.30\\
Soldier&1059081&7.58&4.47&&8.75&9.10&&11.09&1.13&&9.22&3.27\\
\midrule
\hline
\textbf{Average}&&\textbf{8.54}&\textbf{5.40}&&\textbf{10.89}&\textbf{9.09}&&\textbf{10.63}&\textbf{0.81}&&\textbf{9.98}&\textbf{3.35}\\
	\bottomrule\\
\end{tabular}}
\end{center}
\end{table*}

\section{Experimental Results}
\label{performanceeval}

\subsection{Experimental Setup}
\textbf{Training Dataset } To the best of our knowledge, there is no widely used training set for point cloud attribute coding. We consider four different datasets with color attributes, namely MPEG Owlii \cite{d20178i} and 8i \cite{8i} - dynamic dense voxelized full-body point clouds, Microsoft Voxelized Upper Bodies (MVUB) \cite{loop2016microsoft} - dynamic dense voxelized upper-body point clouds, and MPEG CAT1 - static sparse point clouds \cite{8i}. We partition all selected point clouds into a set of points with a maximum bounding box dimension is 64 to reduce the number of sparse training samples.
\\
\textbf{Training Procedure} The training and testing process were performed on a GeForce RTX 3090 GPU with an Adam optimizer. We found that applying an initial learning rate of 5e-4 with a step decay by a factor of 0.75 every 5 epochs gives optimal results. We used a batch size of 128, and an early stopping with the patience of 20 epochs.
\\
\textbf{Test point clouds:} In this work, we aim at dense point cloud coding and follow \cite{huang2021hierarchical}, we evaluate our methods on a set of dense point clouds from the MVUB and MPEG datasets. All selected point clouds are either used in the MPEG Common Test Condition or the JPEG Pleno Common Test Condition to evaluate point cloud coding schemes and have not been used in training.
\subsection{Experimental Results}
Because there are limited works on lossless point cloud attribute compression, especially learning-based methods, we compare our method against two different versions of G-PCC (v14, v11) and the recently proposed method HBDC \cite{huang2021hierarchical}.
Table \ref{tab:compare} reports the encoding bitrate in bits per point (bpp) and encoding time of MNeT, G-PCC, and HBDC with the numbers extracted from the HBDC paper. Generally, we observe that G-PCC v14 still provides the best compression rate.  The introduction of new attribute compression tools (residual coding, attribute prediction scheme, coordinate conversion, etc.) in G-PCC v14 greatly improves the previous version G-PCC v11. HBDC costs slightly fewer bits for color attribute coding than G-PCCv11. Most importantly, the average coding bitrate of our method MNeT is 9.98 bpp, representing 8.4\% and 6.3\% of bitrate reduction compared to G-PCCv11 and HBDC. In terms of encoding time, our encoding time is substantially faster than both G-PCC versions even though we do not deploy any optimization for running time. It should be noted that HBDC and G-PCC are octree-based methods which are generally fast.
\par
\begin{figure}
\captionsetup{singlelinecheck = false, justification=justified, font=small, labelsep=space}
\begin{minipage}[b]{.55\linewidth}
  \centering
  \centerline{\includegraphics[width=0.8\linewidth]{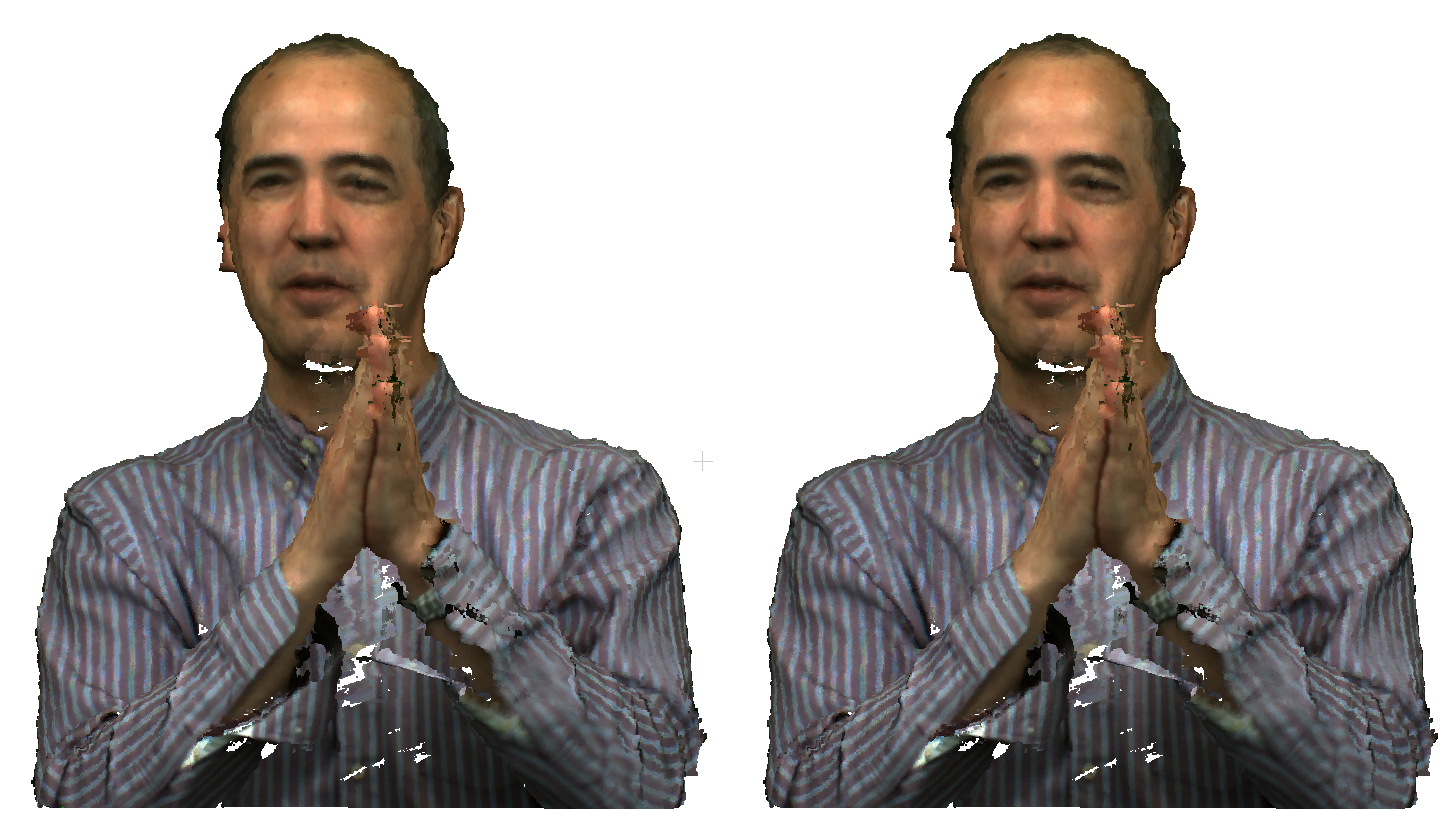}}
  \centerline{(a)}\medskip
\end{minipage}
\hfill
\begin{minipage}[b]{0.40\linewidth}
\label{sfig:typeA}
  \centering
  \centerline{\includegraphics[width=0.9\linewidth]{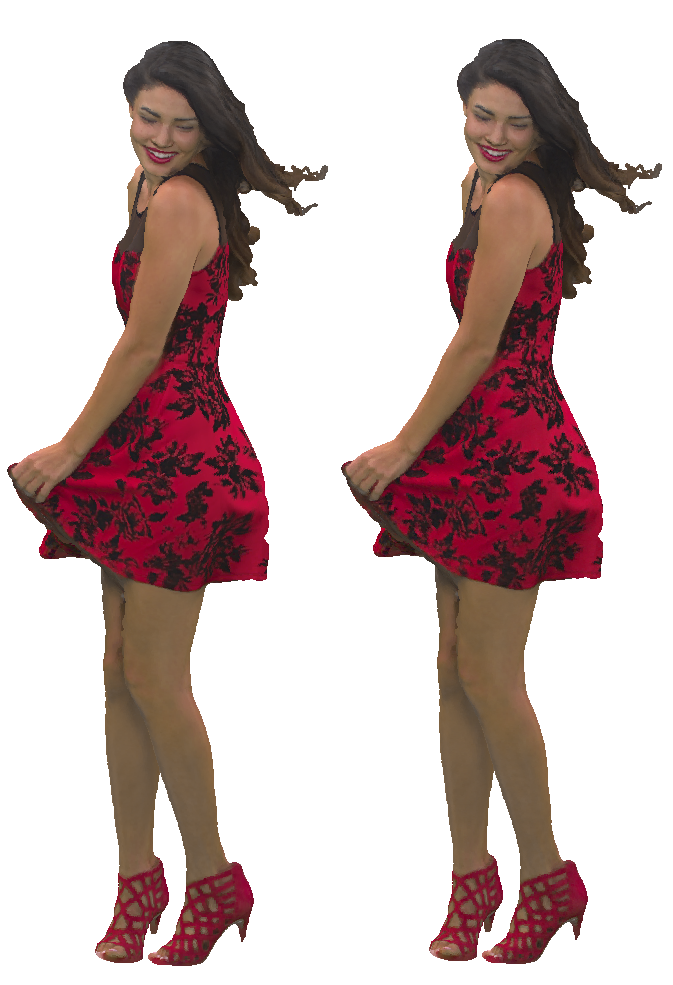} }
  \centerline{(b)}\medskip
\end{minipage}

\hfill

\caption{Lossless point clouds and lossy point clouds sampled from the predicted probability distribution. (a): Lossless Phil (10.49 bpp) and the sampled point cloud (2.83 bpp). (b): Lossless Red\_and\_black (12.23 bpp) and the sampled point cloud (3.30 bpp).}
\label{fig:trainingperformance}
\end{figure}

 Instead of using the probability distributions to losslessly encode the point cloud with an arithmetic coder, we can sample or thresholding the distribution to obtain lossy versions of a point cloud with a much lower bitrate. We show a sampling example at the RGB scale (only extract the bitstream of lower scales) in Figure \ref{fig:trainingperformance}. We do not notice the difference between the lossless and lossy point clouds (less than 27\% bitrate) which shows the efficiency of MNeT in capturing color attributes into latent spaces.

\section{Conclusions}
\label{conclusion}
In this paper, we have presented an end-to-end learning-based lossless point cloud attribute coding method with scalability called MNeT. We built a multiscale architecture to predict the probability distribution of a higher scale based on lower scale representation. This approach enables us to build an accurate context model for arithmetic coding with a single network prediction. The proposed method outperforms G-PCCv11 and HBDC with 8.4\% and 6.3\% bitrate reduction while rivaling the performance of G-PCC v14. In addition, we obtain a substantial improvement in computing efficiency compared G-PCC. We expect further bitrate reduction when training on a more diverse datasets as well as exploiting the neighboring dependencies between points.




\bibliographystyle{IEEEbib}
\bibliography{references/IEEEabrv.bib, references/refs.bib}

\begin{thebibliography}{10}

\bibitem{graziosi2020overview}
D.~Graziosi, O.~Nakagami, S.~Kuma, A.~Zaghetto, T.~Suzuki, and A.~Tabatabai,
\newblock ``An overview of ongoing point cloud compression standardization
  activities: video-based ({V}-{PCC}) and geometry-based ({G}-{PCC}),''
\newblock {\em APSIPA Transactions on Signal and Information Processing}, vol.
  9, 2020.

\bibitem{7025414}
C.~Zhang, D.~Florêncio, and C.~Loop,
\newblock ``Point cloud attribute compression with graph transform,''
\newblock in {\em Proc. IEEE International Conference on Image Processing
  (ICIP)}, 2014, pp. 2066--2070.

\bibitem{7482691}
R.~L. de~Queiroz and P.~A. Chou,
\newblock ``Compression of 3d point clouds using a region-adaptive hierarchical
  transform,''
\newblock {\em IEEE Transactions on Image Processing}, vol. 25, no. 8, pp.
  3947--3956, 2016.

\bibitem{gu20193d}
S.~Gu, J.~Hou, H.~Zeng, H.~Yuan, and K.-K. Ma,
\newblock ``3d point cloud attribute compression using geometry-guided sparse
  representation,''
\newblock {\em IEEE Transactions on Image Processing}, vol. 29, pp. 796--808,
  2019.

\bibitem{choy20194d}
C.~Choy, J.~Gwak, and S.~Savarese,
\newblock ``4d spatio-temporal convnets: Minkowski convolutional neural
  networks,''
\newblock in {\em Proceedings of the IEEE Conference on Computer Vision and
  Pattern Recognition}, 2019, pp. 3075--3084.

\bibitem{TMC13}
M.~Group,
\newblock ``{MPEG} tmc13 reference software,'' (accessed October 25, 2022).

\bibitem{lifting}
K.~Mammou, A.~Tourapis, J.~Kim, F.~Robinet, V.~Valentin, and Y.~Su,
\newblock ``Proposal for improved lossy compression in tmc1,''
\newblock in {\em ISO/IEC JTC1/SC29/WG11 M42640, 2018}.

\bibitem{predicting}
M.~3DG,
\newblock ``{PCC} test model category 3 v0.,''
\newblock in {\em ISO/IEC JTC1/SC29/ WG11 N17249, 2017.}

\bibitem{7434610}
R.~{Mekuria}, K.~{Blom}, and P.~{Cesar},
\newblock ``Design, implementation, and evaluation of a point cloud codec for
  tele-immersive video,''
\newblock {\em IEEE Transactions on Circuits and Systems for Video Technology},
  vol. 27, no. 4, pp. 828--842, 2017.

\bibitem{quach2020folding}
M.~Quach, G.~Valenzise, and F.~Dufaux,
\newblock ``Folding-based compression of point cloud attributes,''
\newblock in {\em Proc. IEEE International Conference on Image Processing
  (ICIP)}. IEEE, 2020, pp. 3309--3313.

\bibitem{9447226}
X.~Sheng, L.~Li, D.~Liu, Z.~Xiong, Z.~Li, and F.~Wu,
\newblock ``Deep-pcac: An end-to-end deep lossy compression framework for point
  cloud attributes,''
\newblock {\em IEEE Transactions on Multimedia}, pp. 1--1, 2021.

\bibitem{wang2022sparse}
J.~Wang and Z.~Ma,
\newblock ``Sparse tensor-based point cloud attribute compression,''
\newblock {\em arXiv preprint arXiv:2204.01023}, 2022.

\bibitem{huang2021hierarchical}
Y.~Huang, B.~Wang, C.-C.~J. Kuo, H.~Yuan, and J.~Peng,
\newblock ``Hierarchical bit-wise differential coding (hbdc) of point cloud
  attributes,''
\newblock in {\em Proc. ICASSP IEEE International Conference on Acoustics,
  Speech and Signal Processing (ICASSP)}. IEEE, 2021, pp. 4215--4219.

\bibitem{yin2021lossless}
Q.~Yin, Q.~Ren, L.~Zhao, W.~Wang, and J.~Chen,
\newblock ``Lossless point cloud attribute compression with normal-based intra
  prediction,''
\newblock in {\em Proc. IEEE International Symposium on Broadband Multimedia
  Systems and Broadcasting (BMSB)}. IEEE, 2021, pp. 1--5.

\bibitem{song2021layer}
F.~Song, Y.~Shao, W.~Gao, H.~Wang, and T.~Li,
\newblock ``Layer-wise geometry aggregation framework for lossless lidar point
  cloud compression,''
\newblock {\em IEEE Transactions on Circuits and Systems for Video Technology},
  vol. 31, no. 12, pp. 4603--4616, 2021.

\bibitem{salimans2017pixelcnn++}
T.~Salimans, A.~Karpathy, X.~Chen, and D.~P. Kingma,
\newblock ``{PixelCNN}++: {Improving} the {PixelCNN} with {Discretized}
  {Logistic} {Mixture} {Likelihood} and {Other} {Modifications},''
\newblock {\em arXiv:1701.05517 [cs, stat]}, January 2017.

\bibitem{nguyen2022learning}
D.~T. Nguyen and A.~Kaup,
\newblock ``Learning-based lossless point cloud geometry coding using sparse
  representations,''
\newblock {\em arXiv preprint arXiv:2204.05043}, 2022.

\bibitem{lim2017enhanced}
B.~Lim, S.~Son, H.~Kim, S.~Nah, and K.~Mu~Lee,
\newblock ``Enhanced deep residual networks for single image
  super-resolution,''
\newblock in {\em Proceedings of the IEEE conference on computer vision and
  pattern recognition workshops}, 2017, pp. 136--144.

\bibitem{d20178i}
E.~d'Eon, B.~Harrison, T.~Myers, and P.~A. Chou,
\newblock ``8i {Voxelized} {Full} {Bodies} - {A} {Voxelized} {Point} {Cloud}
  {Dataset},''
\newblock in {\em {ISO}/{IEC} {JTC1}/{SC29} {Joint} {WG11}/{WG1}
  ({MPEG}/{JPEG}) input document {WG11M40059}/{WG1M74006}}. Geneva, January
  2017.

\bibitem{8i}
``Common test conditions for {PCC},''
\newblock in {\em {ISO}/{IEC} {JTC1}/{SC29}/{WG11} {MPEG} output document
  {N19324}}.

\bibitem{loop2016microsoft}
C.~Loop, Q.~Cai, S.~O. Escolano, and P.~A. Chou,
\newblock ``Microsoft voxelized upper bodies - a voxelized point cloud
  dataset,''
\newblock in {\em {ISO}/{IEC} {JTC1}/{SC29} {Joint} {WG11}/{WG1}
  ({MPEG}/{JPEG}) input document m38673/{M72012}}. May 2016.

\end{thebibliography}

\end{document}